\documentclass[11pt,prd,superscriptaddress,preprintnumbers,amsmath,amssymb,nofootinbib]{revtex4}
\usepackage{epsfig}  
\usepackage{graphicx}
\usepackage{hyperref}
\usepackage{color}
\usepackage{float}
\usepackage{amsfonts}
\usepackage{amsmath}
\usepackage{slashed}
\usepackage{braket}
\usepackage{soul}

\large

\begin{document} 
\title{New physics effects in radiative leptonic $B_s$ decay}

\author{Gauhar Abbas}
\email{gauhar.app@iitbhu.ac.in}
\affiliation{Department of Physics, Indian Institute of Technology (BHU), Varanasi, India}
	
\author{Ashutosh Kumar Alok}
\email{akalok@iitj.ac.in }
\affiliation{Indian Institute of Technology Jodhpur, Jodhpur 342011, India}

\author{Shireen Gangal}
\email{shireen.gangal@gmail.com}
\affiliation{Center for Neutrino Physics, Department of Physics, Virginia Tech, Blacksburg, VA 24061, USA}

\date{\today} 

\preprint{}

\begin{abstract}
There are several anomalous measurements in the decays induced by the quark level transition  $b \to s \mu^+ \mu^-$, which do not agree with the predictions of the Standard Model. These measurements could be considered as signatures of physics beyond the Standard Model. Working within the framework of effective field theory, several groups have performed global fits to all $b \to s \mu^+ \mu^-$ data to identify the Lorentz structure of possible new physics. Many new physics scenarios have been suggested to explain the anomalies in the $b \to s \mu^+ \mu^-$  sector. In this work,  we investigate the impact of these new physics scenarios on the  radiative leptonic decay of  $B_s$ meson. We consider the branching ratio of this decay along with the ratio $R_{\gamma}$  of the differential distribution $B_s \to \mu^+ \mu^- \gamma$ \& $B_s \to e^+ e^- \gamma$ and the muon forward-backward asymmetry $A_{FB}$. We find that the predicted values of the branching ratio and $A_{FB}$ are close to the SM results for all allowed new physics solutions whereas some of the solutions allow large deviation in $R_{\gamma}$ from its SM prediction. 
\end{abstract}

\maketitle 

\newpage
\section{Introduction} 

Over the last few years there have been several measurements in the $B$-meson sector, more specifically in decays induced by the flavor changing neutral current (FCNC) quark level transition  $ b \rightarrow s\, l^+ \, l^-$, which are incompatible with the predictions of the Standard Model (SM).  The following measurements have been of rigorous attention:

\begin{itemize}
\item In 2012, the LHCb collaboration reported the measurement of the ratio $R_K \equiv  \Gamma(B^+ \to K^+ \,\mu^+\,\mu^-)/\Gamma(B^+ \to K^+\,e^+\,e^-)$ performed in the low dilepton invariant mass-squared $q^2$ range ($1.0 \le q^2 \le 6.0 \, {\rm GeV}^2$) \cite{rk}, which deviates from the SM prediction of order $\simeq 1$ \cite{Hiller:2003js, Bordone:2016gaq} by 2.6 $\sigma$. This could be an indication of lepton flavor universality violation in the $b \to s l^+ l^-$ sector.

\item The measurement of $R_K$  was further corraborated recently by the measurement of the $R_{K^*} \equiv \Gamma (B^0 \to K^{*0} \mu^+\mu^-)/\Gamma(B^0 \to K^{*0} e^+ e^-)$ in the low ($0.45 \le q^2 \le 1.1 \, {\rm GeV}^2$)  and central ($1.1 \le q^2 \le 6.0 \, {\rm GeV}^2$) $q^2$ bins \cite{rkstar}.
These measurements  differ from the SM prediction of  order $\simeq 1$ \cite{Hiller:2003js, Bordone:2016gaq} by 2.2-2.4$\sigma$ and 2.4-2.5$\sigma$, in the low and central $q^2$ regions, respectively.  

\item The experimentally measured values of some of the angular observables in $B \to K^* \mu^+ \mu^-$ \cite{Kstarlhcb1,Kstarlhcb2,KstarBelle} disagree with their SM predictions \cite{sm-angular}. In particular, the angular observable $P'_5$ in the  $q^2$-bin 4.3-8.68 disagress with the SM at the level of 4$\sigma$. The recent ATLAS \cite{kstaratlas} and CMS \cite{kstarcms} measurements confirm this disagreement. 

\item  The measured value of the branching ratio of $B_s \to \phi \mu^+ \mu^-$ \cite{bsphilhc1,bsphilhc2} does not agree with its SM value. This disagreement is at the level of 3$\sigma$.
\end{itemize}
The measurements of $R_K$ and $R_{K^*}$ could be an indication of presence of new physics in $b \to s \mu^+ \mu^-$ and/or $b \to s e^+ e^-$ sector where as the discrepancies in $P'_5$ and the branching ratio of $B_s \to \phi \mu^+ \mu^-$ are related to  $b \to s \, \mu^+ \, \mu^-$ sector only. Hence it is quite natural to account for all of these anomalies by assuming new physics only in the $b \to s \mu^+ \mu^-$ sector. A recent global fit \cite{Capdevila:2017bsm} also favours this point of view. 

In order to probe the Lorentz structure of new physics in $b \to s \, \mu^+ \, \mu^-$, several model independent analyses have been performed. It is observed that the present $b \to s \, \mu^+ \, \mu^-$ data can be accommodated by new physics in the form of vector ($V$) and axial-vector operators ($A$)
 \cite{Capdevila:2017bsm, Altmannshofer:2017yso,DAmico:2017mtc,Hiller:2017bzc,Geng:2017svp,Ciuchini:2017mik,Celis:2017doq,Alok:2017sui,Alok:2017jgr}. However there is no unique solution. For e.g., new physics operator $O_9=(\bar{s} \gamma^\mu P_L b)\, (\bar{l} \gamma^\mu l)$ alone as well as a combination of operators $O_9$ and $O_{10}=(\bar{s} \gamma^\mu P_L b)\, (\bar{l} \gamma^\mu \gamma^5 l)$ can both account for  all $b \to s \, \mu^+ \, \mu^-$ anomalous data.  Therefore one needs additional observables to discriminate between various possible solutions, and hence identify the Lorentz structure of new physics in $b \to s \, \mu^+ \, \mu^-$ sector. In this work we consider radiative leptonic decay of $B_s$ meson to explore such a possibility.

The decay $B_s \to \mu^+ \, \mu^-\, \gamma$ has several advantages over its non radiative counterpart $B_s \to \mu^+ \, \mu^-$. In the SM, the decay $B_s \to \mu^+ \, \mu^-$ is chirally suppressed, and hence has a smaller branching ratio. On the other hand, the decay $B_s \to \mu^+ \, \mu^-\, \gamma$ is free from chirality suppression owing to the emission of a photon  in addition to the muon pair. The photon emission, however, suppresses the branching ratio of $B_s \to \mu^+ \, \mu^-\, \gamma$  by a factor of $\alpha_{em}$. The SM prediction of the branching ratio of $B_s \to \mu^+ \, \mu^-\, \gamma$ is of the order $\sim 10^{-8}$ and hence would be helpful in the 
experimental observation. Further, $B_s \to \mu^+ \, \mu^-$  is sensitive to a wider range of new physics operators as compared to that of $B_s \to \mu^+ \, \mu^-$. It is sensitive to the new physics operators $O_9$, $O_{10}$, and also their chirality flipped counterparts $O'_9=(\bar{s} \gamma^\mu P_R b)\, (\bar{l} \gamma^\mu l)$ and $O'_{10}=(\bar{s} \gamma^\mu P_R b)\, (\bar{l} \gamma^\mu \gamma^5 l)$. Hence it is sensitive to all new physics operators which  can provide a possible explanation for present $b \to s \, \mu^+ \, \mu^-$ anomalies.

From the experimental point of view, the observation of $B_s \to \mu^+ \, \mu^-\, \gamma$ decay is a challenging task. The photon in the final state is difficult to detect, in general, the detection efficiency of photon is smaller as compared to that of the charged leptons.  Further, the photon makes the other daughter particles softer which results in smaller reconstruction efficiencies. Hence the observation of the $B_s \to \mu^+ \, \mu^-\, \gamma$ decay seems to be a rather challenging task. However due to the fact that its branching ratio is $\sim 10^{-8}$, this decay mode might not be beyond the reach of the higher runs of the LHC. In ref. \cite{Dettori:2016zff}, a method was suggested for the detection of $B_s \to \mu^+ \, \mu^-\, \gamma$  by making use of the event sample selected for the measurements of the branching ratio of $B_s \to \mu^+ \, \mu^-$. This would  be applicable at Run 2 of the LHC. 

The decay $B_s \to \mu^+ \, \mu^-\, \gamma$ has been studied in  \cite{Aliev:1996ud,Geng:2000fs,Dincer:2001hu,Kruger:2002gf,Melikhov:2004mk,Alok:2006gp,Balakireva:2009kn,Alok:2010zd,Alok:2011gv,Wang:2013rfa,Guadagnoli:2017quo,Kozachuk:2017mdk}.  
In this work we perform a model independent analysis of $B_{s} \to \mu^+ \, \mu^- \, \gamma$ decay by considering all new physics $V$ and  $A$ operators. Apart from the branching ratio, $B(B_{s} \to \mu^+ \, \mu^- \, \gamma)$, we consider the ratio $R_{\gamma} \equiv  \Gamma(B_s \to \mu^+\,\mu^-\, \gamma)/\Gamma(B_s \to e^+\,e^-\, \gamma)$ and the forward backward asymmetry ($A_{FB}$) of muons. We obtain predictions for these observables for the various new physics solutions which provide a good fit to the  
present $b \to s \mu^+ \mu^-$  data.  We intend to identify the new physics interactions which can provide large deviation  in these observables.

The paper is organized as follows. In Section II, we provide theoretical expressions for various observables in  $B_{s} \to \mu^+ \, \mu^- \, \gamma$ decay in the presence of new physics in the form of $V$ and  $A$ operators. The predictions for $B(B_{s} \to \mu^+ \, \mu^- \, \gamma)$, $R_{\gamma}$ and $A_{FB}$ of muons for the existing new physics solutions are presented in Section III. We provide concluding remarks in Section IV.

\section{$B_{s} \to \mu^+ \, \mu^- \, \gamma$ decay }
The effect of new physics in the $b \rightarrow s\, l^+ \, l^- \gamma $ decays can be  most conveniently probed by making use of the effective field theory approach where the new physics contributions are encoded in the Wilson coefficients of the operators of the $b \rightarrow s\, l^+ \, l^-$
effective Hamiltonian. The decay $B_{s} \to l^+ \, l^- \, \gamma$ is induced by the effective Hamiltonian for the quark level transition $b \rightarrow s\, l^+ \, l^-$ ($l=e,\,\mu$), and is given by,
\begin{eqnarray}
\label{b2qll}
&&H_{\rm eff}^{\rm SM} (b\to s\, l^{+}\,l^{-})\, =\, 
{\frac{G_{F}}{\sqrt2}}\, {\frac{\alpha_{\rm em}}{2\pi}}\, 
V_{tb}V^*_{ts}\, 
\left[\,-2i\,m_b\, {\frac{C_{7\gamma}(\mu)}{q^2}}\cdot
\bar s\sigma_{\mu\nu}q^{\nu}\left (1+\gamma_5\right )b
\cdot{\bar l}\gamma^{\mu}l \right.\nonumber\\
&&\left.\qquad\qquad\quad +\, 
C_{9}(\mu)\cdot\bar s \gamma_{\mu}\left (1\, -\,\gamma_5 \right)   b 
\cdot{\bar l}\gamma^{\mu}l \, +\, 
C_{10}(\mu)\cdot\bar s   \gamma_{\mu}\left (1\, -\,\gamma_5 \right) b 
\cdot{\bar l}\gamma^{\mu}\gamma_{5}l \right], 
\end{eqnarray} 
where $G_F$ is the Fermi constant, and $V_{ts}$, $V_{tb}$ are the elements of the Cabbibo-Kobayashi-Maskawa (CKM) matrix. 
The Wilson Coefficients $C_9$ and $C_{10}$ above are associated with the standard short-distance semi-leptonic operators $O_9$ and $O_{10}$ respectively, 
\begin{align}
O_{9} &= (\bar{s} \gamma^\mu P_L b)\, (\bar{l} \gamma^\mu l) \,\,\,\,,\,\,\,\,
O_{10} = (\bar{s} \gamma^\mu P_L b)\, (\bar{l} \gamma^\mu \gamma^5 l)
\label{eq:O9O10}
\end{align}
where $P_{L,R} = (1 \mp \gamma^5)/2$. 

The remaining dominant contribution to this decay emerges from the Wilson Coefficient $C_{7\gamma}$ associated with the magnetic penguin operator, 
$O_{7} = (\bar{s} \sigma_{\mu\nu} q^\nu P_R b )\, (\bar{l} \gamma^\mu l)$. 
The operators $O_{7,9,10}$ present in the effective Hamiltonian contributing to the $B_s^0 \rightarrow \mu^+ \mu^- \gamma$ amplitude can be parameterised in terms of the $B_s \rightarrow \gamma^*$ form factors as follows \cite{Kozachuk:2017mdk},
\begin{eqnarray}
\label{real}
\label{ffs}
\nonumber
\langle
\gamma^* (k,\,\epsilon)|\bar s \gamma_\mu\gamma_5 b|\bar B_s(p)\rangle 
&=& 
i\, e\,\epsilon^*_{\alpha}\, \left ( g_{\mu\alpha} \, k'k-k'_\alpha k_\mu \right )\,\frac{F_A(k'^2,k^2)}{M_{B_s}}, 
\\
\langle \gamma^*(k,\,\epsilon)|\bar s\gamma_\mu b|\bar B_s(p)\rangle
&=& 
e\,\epsilon^*_{\alpha}\,\epsilon_{\mu\alpha \xi \eta}k'_\xi k_\eta\frac{F_V(k'^2,k^2)}{M_{B_s}},   
\\
\langle\gamma^*(k,\,\epsilon)|\bar s \sigma_{\mu\nu}\gamma_5 b|\bar B_s(p) 
\rangle\, k'^{\nu}
&=& 
e\,\epsilon^*_{\alpha}\,\left( g_{\mu\alpha}\,k'k- k'_{\alpha}k_{\mu}\right)\, 
F_{TA}(k'^2, k^2), 
\nonumber
\\
\langle
\gamma^*(k,\,\epsilon)|\bar s \sigma_{\mu\nu} b|\bar B_s(p)\rangle\, k'^{\nu}
&=& 
i\, e\,\epsilon^*_{\alpha}\epsilon_{\mu\alpha \xi \eta} k'_\xi k_\eta F_{TV}(k'^2, k^2)\,,
\nonumber 
\end{eqnarray}
where $k$ is the momentum of the photon emitted from the valence quark of the $B_s$ meson and $k'$ is the momentum emitted from the $b\rightarrow s$ penguin vertex.  

The form factors relevant for this process are, $F_i (k'^2 = q^2, k^2 = 0) = F_i (q^2)$ and $F_i (k'^2=0, k^2 = q^2) = F_i (0,q^2) $, with $q^2$ being the momentum of the lepton pair and $i = \{V, A, TV, TA \}$. In this work we consider  both single and double pole parametrization of these form factors. The form factors $F_i(q^2)$ in the single pole parametrization are given by,
\begin{equation}
\label{modifiedpole-s}
F_i(q^2)= \beta_i \frac{f_{Bs} m_{Bs}}{\Delta_i + 0.5\, m_{Bs}(1-q^2/m_{Bs}^2) }
\,,
\end{equation}
where the parameters $f_{Bs}$, $\beta_i$ and $\Delta_i$ can be found in Ref.~\cite{Kruger:2002gf}. The form factors in the double pole parametrization use a modified pole parametrization to explicitly take into account the poles at $q^2 = M_R^2$ where $M_R$ is the mass of the meson and provide better precision in the entire region of $q^2$.  These form factors are parameterized as,
\begin{equation}
\label{modifiedpole-d}
F_i(q^2)=\frac{F_i(0)}{(1-{q^2}/{M_{R_i}^2})(1-\sigma_1({q^2}/{M_{R_i}^2})+\sigma_2({q^2}/{M_{R_i}^2})^2)}\,.
\end{equation}
The details of the calculation of these form factors as well as the numerical values of the parameters can be found in Ref.~\cite{Kozachuk:2017mdk}.
Further,  the form factors $F_i(0, q^2)$ for $i = TV, TA$ are given by,  
\begin{eqnarray}
\label{vmd}
F_{TV,TA}(0, q^2) = F_{TV,TA}(0, 0)\, -\,\sum_V\,2\,f_V^{\rm e.m.} g^{B\to V}_+(0)\frac{q^2/M_V}{q^2\, -\, M^2_V\, +\, iM_V\Gamma_V},
\end{eqnarray}
where the values of the mass and width of the vector meson resonances,  $M_V$ and $\Gamma_V$, respectively and the  $B\to V$ transition form factors $g^{B\to V}_+(0)$ can be found in Ref.~\cite{Kozachuk:2017mdk}. 

The two form factor parametrizations agree upto 10\% in the low $q^2$ region below 15 $\mathrm{GeV}^2$ and deviate from each other by upto 20\% in the high $q^2$ region. In our analyses we consider systematic uncertainty arising from the form factors to be about 10\% for both parameterizations.

The global analyses of the $b \rightarrow s l^+ l^-$ anomalies have shown that if new physics is present, it will contribute mainly via the operators $O_9$ and $O_{10}$ and their chirality flipped counterparts. Therefore to study new physics effects in the $B_{s} \rightarrow l^+ l^- \gamma $ transition, we consider new physics in the form of 
$O^{(')}_9$ and  $O^{(')}_{10}$ operators.  The effective Hamiltonian in the presence of these additional new physics operators is,
\begin{eqnarray}
H_{\mathrm{eff}}(b\to s \, l^{+} \, l^{-}) = H^{\rm SM}_{\mathrm{eff}} (b\to s \, l^{+} \, l^{-}) + H^{\rm VA}_{\mathrm{eff}} (b\to s \, l^{+} \, l^{-}),
\end{eqnarray}
where $H^{\rm VA}_{\mathrm{eff}}(b\to s \, l^{+} \, l^{-})$  is given by, 
\begin{align} \nonumber
H^{\rm VA}_{\mathrm{eff}}(b\to s \, l^{+} \, l^{-}) &=  \frac{\alpha G_F}{\sqrt{2} \pi} V_{ts}^* V_{tb} \bigg[C_9^{NP} (\overline{s} \gamma^{\mu} P_L b)(\overline{l} \gamma_{\mu} l) + C_{10}^{NP} (\overline{s} \gamma^{\mu} P_L b)(\overline{l} \gamma_{\mu} \gamma_{5} l)  \\
 &~~~~~~~~~~~~~~~ + C_9'^{NP} (\overline{s} \gamma^{\mu} P_R b)(\overline{l} \gamma_{\mu} l) + C_{10 }'^{NP} (\overline{s} \gamma^{\mu} P_R b)(\overline{l} \gamma_{\mu} \gamma_{5} l) \bigg],
 \label{eq:NPHeff}  
\end{align}
where $C_9^{NP}$ and $C_{10}^{NP} $ are the new physics couplings associated with the operators $O_{9}$ and $O_{10}$, respectively while $C_9'^{NP}$ and $C_{10}'^{NP} $ are the coefficients of the 
 the primed operators $O_{9}^{'}$ and $O_{10}^{'}$, respectively which are obtained by replacing $P_L$ by $P_R$ in Eq.~\ref{eq:O9O10}.

The decay $B_s^0 \rightarrow \mu^+ \mu^- \gamma$ receives contributions from many channels \cite{Kozachuk:2017mdk, Guadagnoli:2017quo}. We present the amplitudes for these channels in the presence of additional new physics VA operators, defined in Eq.~\ref{eq:NPHeff}\, as follows: 
\begin{itemize}
\item The amplitude for the emission of a real photon from the valence quarks of $B_s$ meson and a lepton pair from the FCNC vertex is given by,
\begin{eqnarray}
\label{A1}
A^{(1)}&=& \langle\gamma (k,\,\epsilon),\,l^+(p_1),\,l^-(p_2)|H_{\rm eff}^{b\to s l^+l^-}|\bar B_s(p) \rangle\, =\,
\frac{G_F}{\sqrt{2}}\, V_{tb}V^*_{ts}\,\frac{\alpha_{\rm em}}{2\pi}\, e\, 
\epsilon^*_{\alpha}
\nonumber \\
&& \times \left[
\Big((C_{9}^{\mathrm{eff}} + C_9^{NP} + C_9'^{NP})P^{\perp}_{\mu \alpha} \frac{F_{V}(q^2)}{M_{Bs}} - (C_{9}^{\mathrm{eff}} + C_9^{NP} - C_9'^{NP})P^{\parallel}_{\mu \alpha}\frac{F_{A}(q^2)}{M_{Bs}} \Big) \bar l (p_2)\gamma_{\mu} l (-p_1)\, \right.
\nonumber \\
&& \left. +\, \Big((C_{10} + C_{10}^{NP} + C_{10}'^{NP})P^{\perp}_{\mu \alpha} \frac{F_{V}(q^2)}{M_{Bs}}- (C_{10} + C_{10}^{NP} - C_{10}'^{NP}) P^{\parallel}_{\mu \alpha}\frac{F_{A}(q^2)}{M_{Bs}} \Big) \bar l (p_2)\gamma_{\mu} \gamma_{5} l (-p_1)\right.
\nonumber \\
&& \left. +\, \frac{2C_{7\gamma}(\mu)}{q^2}m_b \Big(P^{\perp}_{\mu \alpha} F_{TV}(q^2,0) - P^{\parallel}_{\mu \alpha} F_{TA}(q^2,0) \Big)
\bar l (p_2)\gamma_{\mu}l (-p_1)\, 
\right], 
\end{eqnarray}
where
\begin{align}
P^{\perp}_{\mu \alpha} &= \epsilon_{\mu\alpha \xi \eta} k'_\xi k_\eta \,\,\,,\,\,\, P^{\parallel}_{\mu \alpha} = i \left (g_{\mu\alpha}\, k'k\, -\, k'_{\alpha}k_{\mu}\right)\,.
\end{align}
In this process, the momentum from the FCNC vertex, $k' = q$ and the momentum emitted from the valence quark, $k = p-q$. So $k'^2 = q^2$ and $k^2 = 0$ and the form factors $F_{i}(q^2,0)$ given in Eq.~\eqref{modifiedpole-d} contribute.

\item The amplitude for the  emission of a virtual photon from the valence quark of $B_s$ meson and a real photon from the FCNC vertex is,
\begin{eqnarray}
\label{A2}
A^{(2)}&=&\langle\gamma (k',\,\epsilon),\, l^+(p_1),\,l^-(p_2)\left |H_{\rm eff}^{b\to s\gamma} \right|\bar B_s(p) \rangle\,=
\frac{G_F}{\sqrt{2}}\,V_{tb}V^*_{ts} \frac{\alpha_{\rm em}}{2\pi}\, e\,\epsilon^*_{\mu}\bar l (p_2)\gamma_{\alpha} l (-p_1)
\nonumber \\
&\times&\left[
\frac{2 m_b C_{7\gamma}(\mu)}{q^2} \Big(P^\perp_{\mu \alpha}\, F_{TV}(0,q^2) - P^\parallel_{\mu \alpha} F_{TA}(0,q^2)\Big)
\right]\,,
\end{eqnarray}
where  $k^2= q^2$ and $k'^2 =0$ and the form factors appearing in the amplitude are $F_{TV}(0,q^2)$ and $F_{TA}(0, q^2)$ defined in Eq.~\eqref{vmd}.

\item The amplitude for bremsstrahlung emission from the leptons in the final state is given by,
\begin{align}
\label{bremsstrahlung}
A^{\rm Brems}&= -i\, e\,\frac{G_F }{\sqrt{2}}\,\frac{\alpha_{\rm em}}{2\pi}\, V^*_{td}V_{tb}\, 
\frac{f_{B_s}}{M_{B_s}}\, \frac{2\, m_{l}}{M_{Bs}}\,
\bar l (p_2)
\left (
\frac{(\gamma\epsilon^*)\,(\gamma p)}{\hat t-\hat m^2_{l}}\, -\, 
\frac{(\gamma p)\,(\gamma\epsilon^*)}{\hat u-\hat m^2_{l}}
\right )
\gamma_5\, l (-p_1)
\nonumber \\ 
& \,\,\,\, \times \left( C_{10}(\mu) + C_{10}^{NP}- C_{10}'^{NP}\right)\, 
\end{align}
This contribution is suppressed compared to $A^{(1)}$ by the lepton mass but is important in the high $q^2$ region, when $q^2$ approaches $M_{Bs}^2$. 
\end{itemize}

In an effective theory approach, the heavy degrees of freedom like the top quark and W/Z boson masses are integrated out, however the effects of lighter degrees of freedom like charm and up quarks need to be taken into account in the loops. The contribution of the charm quarks to the $B \rightarrow \gamma^* \gamma^*$ amplitude, at the lowest order arise from the charming penguin topology and the weak annihilation topology.   

We now discuss the contribution of the penguin diagrams containing the charm quarks in the loop. The amplitudes of these penguin diagrams can be found in Ref.~\cite{Kozachuk:2017mdk}. These amplitudes have the same Lorentz structure as that of the amplitudes  $A^{(1)}$ and $A^{(2)}$ defined in Eq.~\eqref{A1} and Eq.~\eqref{A2}. Hence, the form factor in these penguin diagram amplitudes can be expressed as corrections to the Wilson coefficient $C_{9}$ appearing in the amplitudes $A^{(1)}$ and $A^{(2)}$. These corrections arising due to the charm loop effects consist of factorizable contribution and non-factorizable contribution arising due to soft gluon exchanges. 
The non-factorizable corrections due to charm-loop effects for the $B\to\gamma l^+l^-$ amplitude have not been calculated yet. These non-factorizable corrections have been computed for $B\to K^* l^+l^-$ amplitude \cite{Khodjamirian:2010vf}, and are a good approximation for $B \rightarrow \mu^+ \mu^- \gamma$  at low $q^2$. We implement the factorizable and non-factorizable corrections in the low $q^2$ region by adding to $C_9$, a simplified $q^2$-dependent parameterization of these corrections, taken from Ref.~\cite{Khodjamirian:2010vf}.

The amplitude of the weak annihilation diagrams including the QCD corrections is \cite{Kozachuk:2017mdk},
\begin{eqnarray}
A^{WA}=-\frac{G_F}{\sqrt{2}}\alpha_{\rm em}e\, a_1
\{V_{ub}V^*_{ud}+V_{cb}V^*_{cd}\}  
\frac{16}{3}
\epsilon_{\mu \varepsilon^* q k}\frac{1}{q^2}\,\bar l \gamma_\mu l.  
\label{eq:WA}
\end{eqnarray}

The amplitude $A^{(2)}$ has a structure similar to the  $C_{7\gamma}$ part of $A^{(1)}$ except for the form factors. Similarly, the Lorentz structure of the weak annihilation amplitude as given in Eq.~\eqref{eq:WA} is the same as that of $C_{7\gamma} P_{\mu \alpha}^\perp$ in the amplitude $A^{(1)}$. These two amplitudes can therefore be combined with $A^{(1)}$ by redefining the form factors as follows,  
\begin{align}
\bar{F}_{TV}(q^2) &= F_{TV}(q^2,0) + F_{TV}(0, q^2)- \frac{16}{3} \frac{V_{ub}V_{us}^* + V_{cb}V_{cs}^*}{V_{tb}V_{ts}^*}\frac{a_1\,f_b}{C_{7\gamma}\,m_b} \\
\bar{F}_{TA}(q^2) &= F_{TA}(q^2,0) + F_{TA}(0, q^2) 
\end{align}

The double differential decay rate for $B_s \rightarrow \mu^+ \mu^- \gamma$ process in the presence of  new physics V and A operators can now be calculated using the amplitudes and form factors defined above. It is expressed as a sum of three contributions: 
\begin{itemize}
\item ${d^2\Gamma^{(1)}}/{d\hat s\, d\hat t}$, which receives contributions from  the combined amplitudes ($A^{(1)} +  A^{(2)}+ A^{(WA)}$),
\item ${d^2\Gamma^{(2)}}/{d\hat s\, d\hat t}$,  which is the contribution from the bremsstrahlung amplitude $A^{\mathrm{Brems}}$, and 
\item ${d^2\Gamma^{(12)}}/{d\hat s\, d\hat t}$,  which is the mixing between the amplitudes $A^{(1 + 2 + WA)}$ and $A^{\mathrm{Brems}}$. 
\end{itemize}
These decay rates are as given below,
\begin{align}
\label{Gamma1}
\frac{d^2\Gamma^{(1)}}{d\hat s\, d\hat t}\, &=\, 
\frac{G^2_F\,\alpha^3_{em}\, M^5_{Bs}}{2^{10}\,\pi^4}\, 
\left |V_{tb}\, V^*_{ts} \right |^2
\left [ 
x^2\, 
B_0\left (\hat s,\,\hat t\right )\, +\,
x\,\,\xi\left (\hat s,\hat t\right )\,\tilde B_1\left (\hat s,\,\hat t\right )
\, +\,  
\xi^2\left (\hat s,\hat t\right )\,\tilde B_2\left (\hat s,\,\hat t\right ) 
\right ], 
\end{align}
where
\begin{align}
B_0\left (\hat s,\,\hat t\right )\, &=\,
    \left (\hat s\, +\, 4\hat m^2_{l} \right )
    \left (F_1\left(\hat s\right )\, +\, F_2\left(\hat s\right )\right)\, -\, 
    8\hat m^2_{l}\,\left |C_{10}(\mu) + C_{10}^{NP} + C_{10}'^{NP} \right |^2
    F^2_V(q^2 )\, + \nonumber\\
     & \left |C_{10}(\mu) + C_{10}^{NP} - C_{10}'^{NP} \right |^2 F^2_A (q^2 ), 
    \nonumber \\
B_1\left (\hat s,\,\hat t\right )\, &=\,
     8\,[
                \hat s\, F_V(q^2)\, F_A(q^2)\, 
                Re \left[\left(C^{\rm eff\, *}_{9}(\mu, q^2)+ C_9^{*NP}\,+ C_9'^{*NP}\right)\, \left(C_{10}(\mu)+C_{10}^{NP}- C_{10}'^{NP} \right)\right]\, 
         \nonumber\\
&  +\,  
                \hat m_b\, F_V(q^2)\, Re \left[ C^*_{7\gamma}(\mu)\, \bar F^*_{TA}(q^2)\, \left(C_{10}(\mu)+ C_{10}^{NP}-C_{10}'^{NP}\right) \right]
\nonumber \\
& +\,
         \hat m_b\, F_A(q^2)\, Re \left[C^*_{7\gamma}(\mu)\, \bar F^*_{TV}(q^2)\, \left(C_{10}(\mu)+ C_{10}^{NP}-C_{10}'^{NP}\right) \right] 
         ,\nonumber \\
B_2\left (\hat s,\,\hat t\right )\, &=\,\hat s\, 
    \left (F_1\left(\hat s\right )\, +\, F_2\left(\hat s\right )\right), 
\\
\text{and} \nonumber \\ 
F_1\left (\hat s\right )\, &=\, 
   \left (\left |C^{\rm eff}_{9}(\mu, q^2)+ C_9^{NP} + C_9'^{NP} \right |^2 +
   \left |C_{10}(\mu) + C_{10}^{NP} + C_{10}'^{NP} \right |^2  \right)F^2_V(q^2)
    + 
   \left (\frac{2\hat m_b}{\hat s}\right )^2 \nonumber\\ 
   & \left |C_{7\gamma}(\mu)\, \bar F_{TV}(q^2)\right |^2 + \frac{4\hat m_b}{\hat s}\, F_V(q^2)\, 
   Re\left[ C_{7\gamma}(\mu)\, \bar F_{TV}(q^2)\, \left(C^{\rm eff\, *}_{9}(\mu, q^2)+ C_9^{*NP} + C_9'^{*NP} \right) 
     \right ] \nonumber\\
     F_2\left (\hat s\right )\, &=\, 
   \left (\left |C^{\rm eff}_{9}(\mu, q^2)+ C_9^{NP} - C_9'^{NP} \right |^2 +
   \left |C_{10}(\mu) + C_{10}^{NP} - C_{10}'^{NP} \right |^2  \right)F^2_A(q^2)
    + 
   \left (\frac{2\hat m_b}{\hat s}\right )^2 \nonumber\\ 
   & \left |C_{7\gamma}(\mu)\, \bar F_{TA}(q^2)\right |^2 + \frac{4\hat m_b}{\hat s}\, F_A(q^2)\, 
   Re\left[ C_{7\gamma}(\mu)\, \bar F_{TA}(q^2)\, \left(C^{\rm eff\, *}_{9}(\mu, q^2)+ C_9^{*NP} - C_9'^{*NP} \right) 
     \right ] 
\end{align}

\begin{align}
\label{Gamma2}
\frac{d^2\Gamma^{(2)}}{d\hat s\, d\hat t} &=  
\frac{G^2_F\,\alpha^3_{em}\, M^5_{Bs}}{2^{10}\,\pi^4}\, 
\left |V_{tb}\, V^*_{tq} \right |^2\,
\left (\frac{8\, f_{B_s}}{M_B}\right )^2\,\hat m^2_{l}\,
\left |C_{10A}(\mu) + C_{10}^{NP} - C_{10}'^{NP} \right |^2 \times
\nonumber \\
&  \,\,\,\left [
   \frac{\hat s\, +\, x^2/2}
         {(\hat u\, -\,\hat m^2_{l})(\hat t\, -\,\hat m^2_{l})}\, 
-\,\left (\frac{x\,\hat m_{l}}
        {(\hat u\, -\,\hat m^2_{l})\, (\hat t\, -\,\hat m^2_{l})}
   \right )^2\, 
   \right ]
\\ \\
\label{Gamma12}
\frac{d^2\Gamma^{(12)}}{d\hat s\, d\hat t} &=
-\frac{G^2_F\,\alpha^3_{em}\, M^5_{Bs}}{2^{10}\,\pi^4}\, 
\left |V_{tb}\, V^*_{ts} \right |^2\,\frac{16\, f_{B_s}}{M_{Bs}}\, \hat m^2_{l}
\,\frac{x^2}{(\hat u\, -\,\hat m^2_{l})(\hat t\, -\,\hat m^2_{l})}
\Bigg[\frac{2\, x\, \hat m_b}{\hat s}\, Re \Big[(C^*_{10A}(\mu)+ C_{10}^{*NP} - C_{10}'^{* NP})
\nonumber \\
& \,\,\, C_{7\gamma}(\mu)\bar F_{TV}(q^2, 0)\Big] \,
 +\, x\, F_V(q^2)\, Re \Big[(C^*_{10A}(\mu)+C_{10}^{*NP} - C_{10}'^{*NP})(C^{\rm eff}_{9V}(\mu, q^2)+ C_9^{NP} + C_9'^{NP}) \Big]
 \nonumber\\
& +\,\xi(\hat s,\hat t)\, F_A(q^2)\,\left |C_{10A}(\mu)+ C_{10}^{NP} - C_{10}'^{NP} \right |^2 
\Bigg]. 
\end{align}
Here
\begin{eqnarray}
\label{mandelstam}
\hat s\, =\,\frac{\left (p\, -\, k\right )^2}{M_{Bs}^2},\quad 
\hat t\, =\,\frac{\left (p\, -\, p_1\right )^2}{M_{Bs}^2},\quad 
\hat u\, =\,\frac{\left (p\, -\, p_2\right )^2}{M_{Bs}^2}, 
\end{eqnarray} 
with 
$\hat s\, +\,\hat t\, +\,\hat u\, =\, 1\, +\, 2\hat m^2_{l}$, 
$\hat m^2_{l}\, =\,  m^2_{l}/M^2_{Bs}$, 
$\hat m_b\, =\, m_b/M_{Bs}$, and 
\cite{Kruger:2002gf} 
\begin{eqnarray}
x\, =\, 1\, -\, \hat s,\qquad 
\cos\theta\, =\,\frac{\xi\left (\hat s,\hat t\right )}
          {x\,\sqrt{1\, -\, 4\hat m^2_l/\hat s}},\qquad 
\xi\left (\hat s,\hat t\right )\, =\,\hat u\, -\,\hat t.
\label{eq:costheta}
\end{eqnarray} 

The total differential branching ratio is then given by,
\begin{equation}
\frac{dB (B_s \to \mu^+ \, \mu^-\,\gamma)}{d q^2}= \frac{\tau_{Bs}}{m_{Bs}^2} \int dt \Big(\frac{d^2\Gamma^{(1)}}{d\hat s\, d\hat t} + \frac{d^2\Gamma^{(2)}}{d\hat s\, d\hat t} + \frac{d^2\Gamma^{(12)}}{d\hat s\, d\hat t} \Big).
\end{equation}

We also consider the ratio,
\begin{equation}
R_{\gamma}(q^2) \equiv \frac{d\Gamma(B_s \to \mu^+ \, \mu^-\,\gamma)/d q^2}{d\Gamma (B_s \to e^+ \, e^-\,\gamma)/d q^2},
\end{equation}
along with the forward backward asymmetry of muons,
\begin{eqnarray}
A_{FB}(q^2)\,=\,\frac{\int\limits_0^1 d\cos\theta \, \frac{d^2\Gamma(B_{s}\to l^+l^-\gamma)}{dq^2 \, 
d\cos\theta}-\int\limits_{-1}^0 d\cos\theta \, \frac{d^2\Gamma(B_{s}\to l^+ l^-\gamma)}{dq^2 \, 
d\cos\theta}}{\frac{d\Gamma(B_{s}\to l^+l^-\gamma)}{dq^2}},
\end{eqnarray}
where $\theta$, the angle between the momentum of $B_{s}$ meson $\vec{p}$ and $\vec{p_2}$, the momentum of the 
lepton, can be expressed in terms of $\hat{t}$ as given in Eq.~\ref{eq:costheta}.

In the next section, we obtain predictions for these observables for various new physics solutions which provide a good fit to the present $b \to s \mu^+ \mu^-$ data.

\section{Results and Discussions}

After the measurement of $R_{K^*}$, several groups have performed global fits to all the $b \to s \, \mu^+ \, \mu^-$ data to identify one or combinations of Wilson coefficients which provide a good fit to the data \cite{Capdevila:2017bsm, Altmannshofer:2017yso,DAmico:2017mtc,Hiller:2017bzc,Geng:2017svp,Ciuchini:2017mik,Celis:2017doq,Alok:2017sui}.  In most of the analyses, three scenarios: (I) $C_9^{\mu \mu}$ (NP) $<0$, (II) $C_9^{\mu \mu}$ (NP) = - $C_{10}^{\mu \mu}$ (NP) $<0$ and (III) $C_9^{\mu \mu}$ (NP) = - $C_{9}^{'\mu \mu}$ (NP) $<0$  were suggested as an explanation of anomalies in the $b \to s \, \mu^+ \, \mu^-$ sector \cite{Capdevila:2017bsm}. The numerical values of the Wilson coefficients corresponding to these scenarios are listed in Table~\ref{table-wc}.
\begin{table}[h]
  \begin{center}
\begin{tabular}{|c|c|c|}
\hline
 Scenario & WC & Operator \\ 
 \hline
(I) $C_9^{\mu \mu}$ (NP) & $-1.25 \pm 0.19$   & $[\bar{s}\gamma_{\mu}P_Lb] \,[\bar{\mu}\gamma^{\mu}\mu]$\\
  \hline
  (II) $C_9^{\mu \mu}$ (NP) = - $C_{10}^{\mu \mu}$ (NP) & $-0.68 \pm 0.12$ & $[\bar{s}\gamma_{\mu}P_Lb] \,[\bar{\mu}\gamma^{\mu}P_L\mu]$\\
   \hline
 (III) $C_9^{\mu \mu}$ (NP) = - $C_{9}^{'\mu \mu}$ (NP) &  $-1.11  \pm 0.17$ &  $[\bar{s}\gamma_{\mu}\gamma_5b] \,[\bar{\mu}\gamma^{\mu}\mu]$ \\
    \hline
    \end{tabular}
        \caption{New physics scenarios suggested as an explanation for all $b \to s \, \mu^+ \, \mu^-$ data. The numerical values of  Wilson coefficients are taken from \cite{Alok:2017sui}.}
            \label{table-wc}
  \end{center}
\end{table}

Eventually these scenarios must arise in some new physics models.  The simplest new physics models that can give rise to these scenarios involve the tree-level exchange of a leptoquark or a $Z'$ boson.  There are three leptoquark models that can explain the data in the $b \to s \, \mu^+ \, \mu^-$ sector. They are scalar triplet with $Y=1/3$ ($S_3$), vector isosinglet with $Y=-2/3$ ($S_3$) and vector isotriplet with $Y=-2/3$ ($U_3$). All of these leptoquark models give rise to scenario (II). The first and third scenarios can only be achieved in $Z'$ models. $Z'$ couples vectorially to $\bar{s}b$ in scenarios (I) and (II), and hence one can easily construct new physics models.
Scenario (III)  requires an axial-vector coupling of the $Z'$ to $\bar{s}b$, and hence it can only arise
in contrived $Z'$ models. Further, scenario (III) predicts $R_K\sim 1$ at the best fit point and hence is in disagreement with the measurement. Therefore  scenario (III) is disfavoured both theoretically and experimentally \cite{Alok:2017sui}. 

We obtain predictions for several observables in the $B_s \to \mu^+ \mu^- \gamma$ decay such as the branching ratio, the ratio $R_{\gamma}$  of the differential distribution $B_s \to \mu^+ \mu^- \gamma$ and $B_s \to e^+ e^- \gamma$ and the muon forward-backward asymmetry $A_{FB}$ for  new physics scenarios listed in Table.~\ref{table-wc}. We look for large deviations in these observables as compared to their SM values. Furthermore, we study the discriminating capability of these observables for various new physics solutions, in particular scenarios (I) and (II). However, for completeness, we also include scenario (III) in our analysis.

\begin{figure}[t] 
\centering
\begin{tabular}{cc}
\includegraphics[width=82mm]{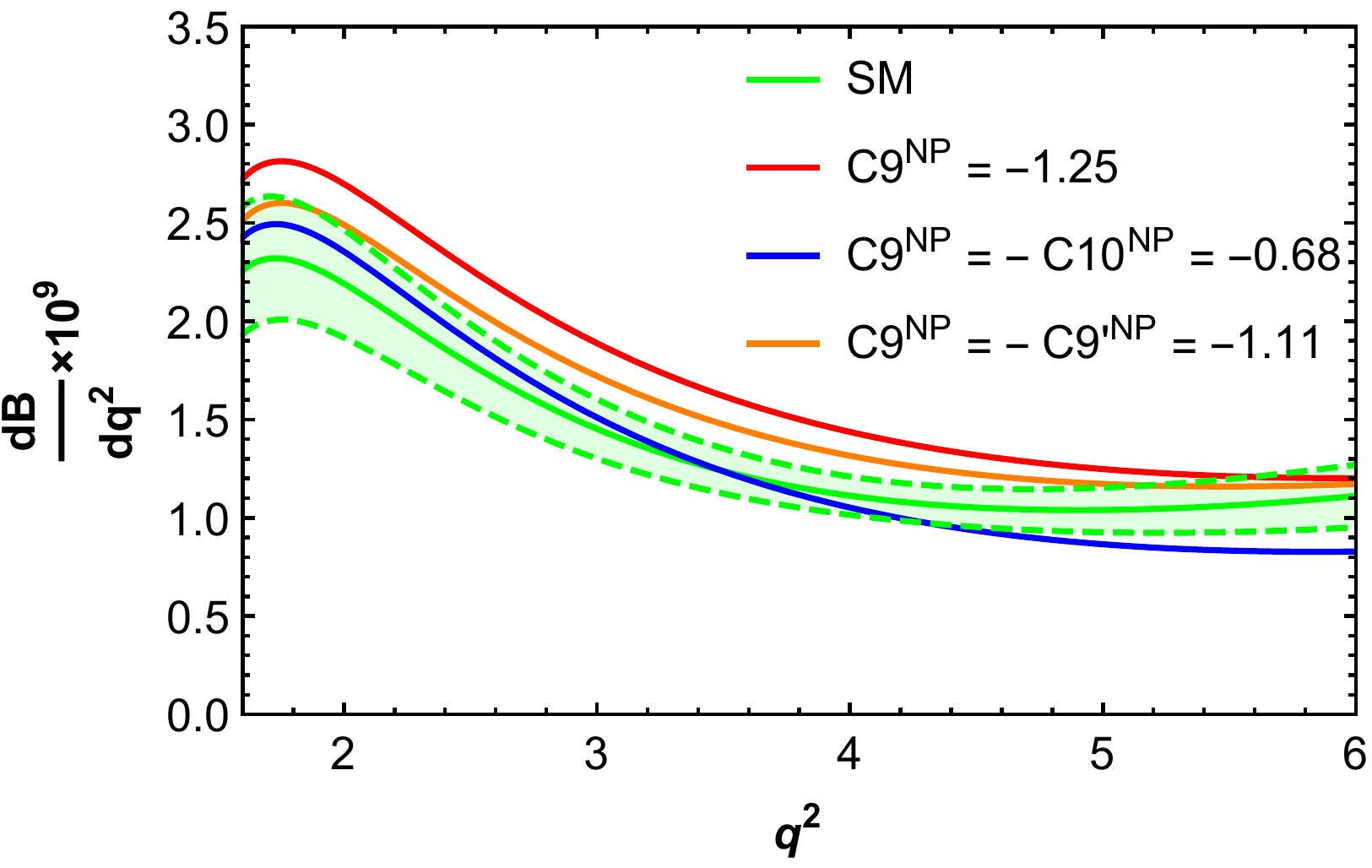}&
\includegraphics[width=82mm]{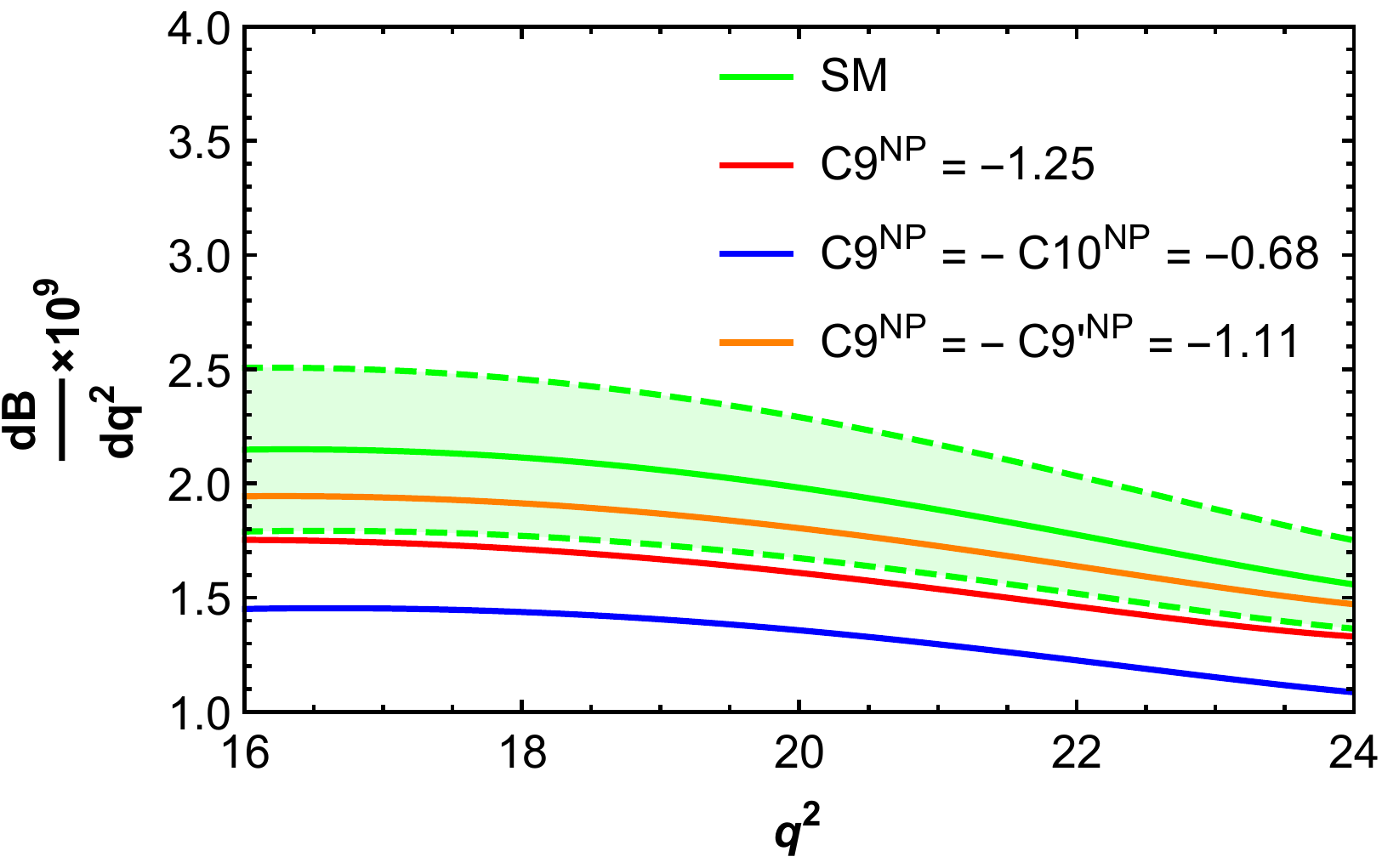}\\
\end{tabular}
\caption{Left and right panels depict the differential branching ratio, $dB/dq^2$, in low (2-6 $\mathrm{GeV}^2$) and high-$q^2$ (15.8-23 $\mathrm{GeV}^2$) regions, respectively for the single pole parameterization of the form factors. The green band corresponds to the 1$\sigma$ range of the SM prediction.  The red, blue and orange curves correspond to $dB/dq^2$ for scenarios (I), (II) and (III), respectively at the best fit values of the new physics WCs.  }
\label{fig-dbr1}
\end{figure}

\begin{figure}[t] 
\centering
\begin{tabular}{cc}
\includegraphics[width=82mm]{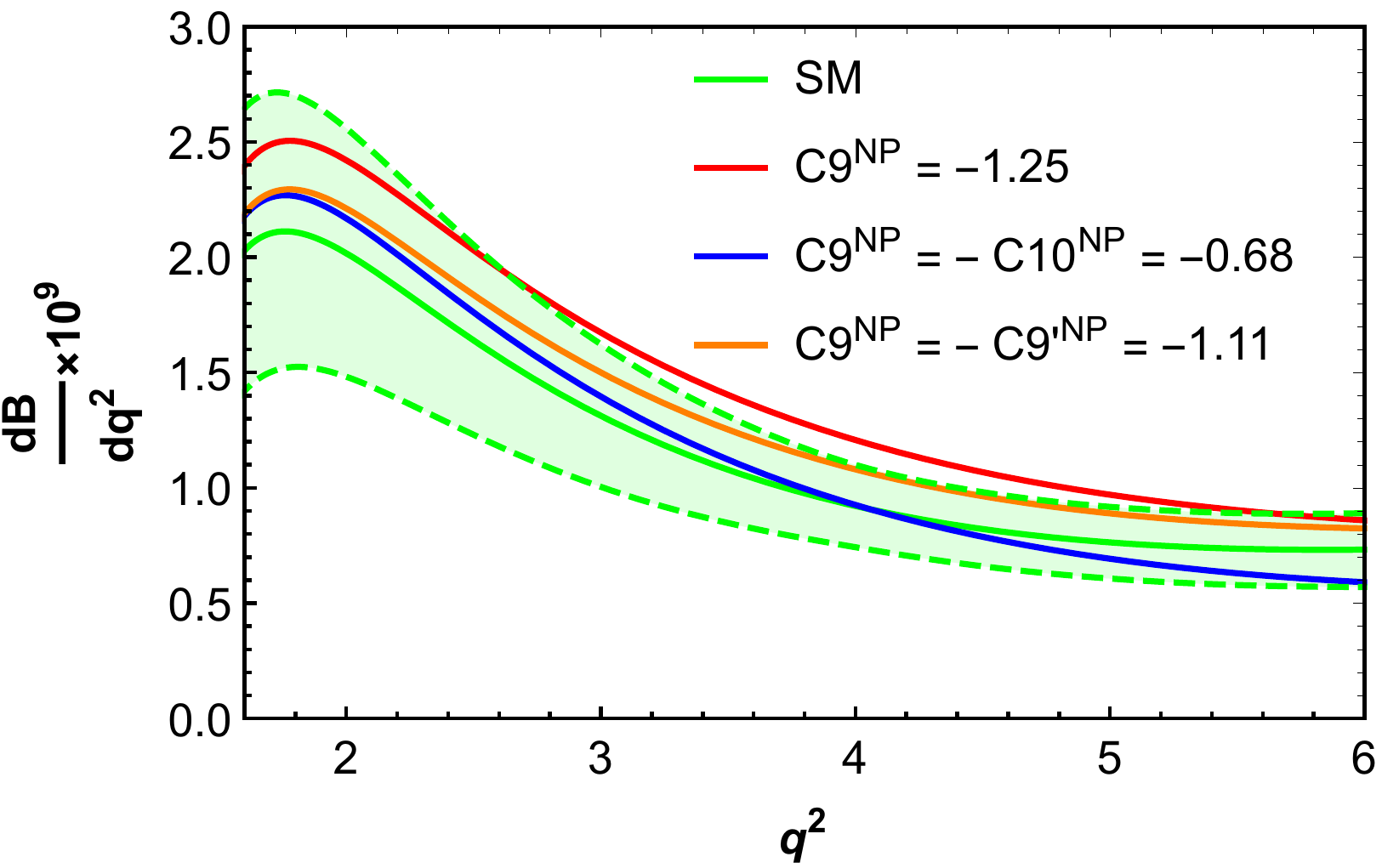}&
\includegraphics[width=82mm]{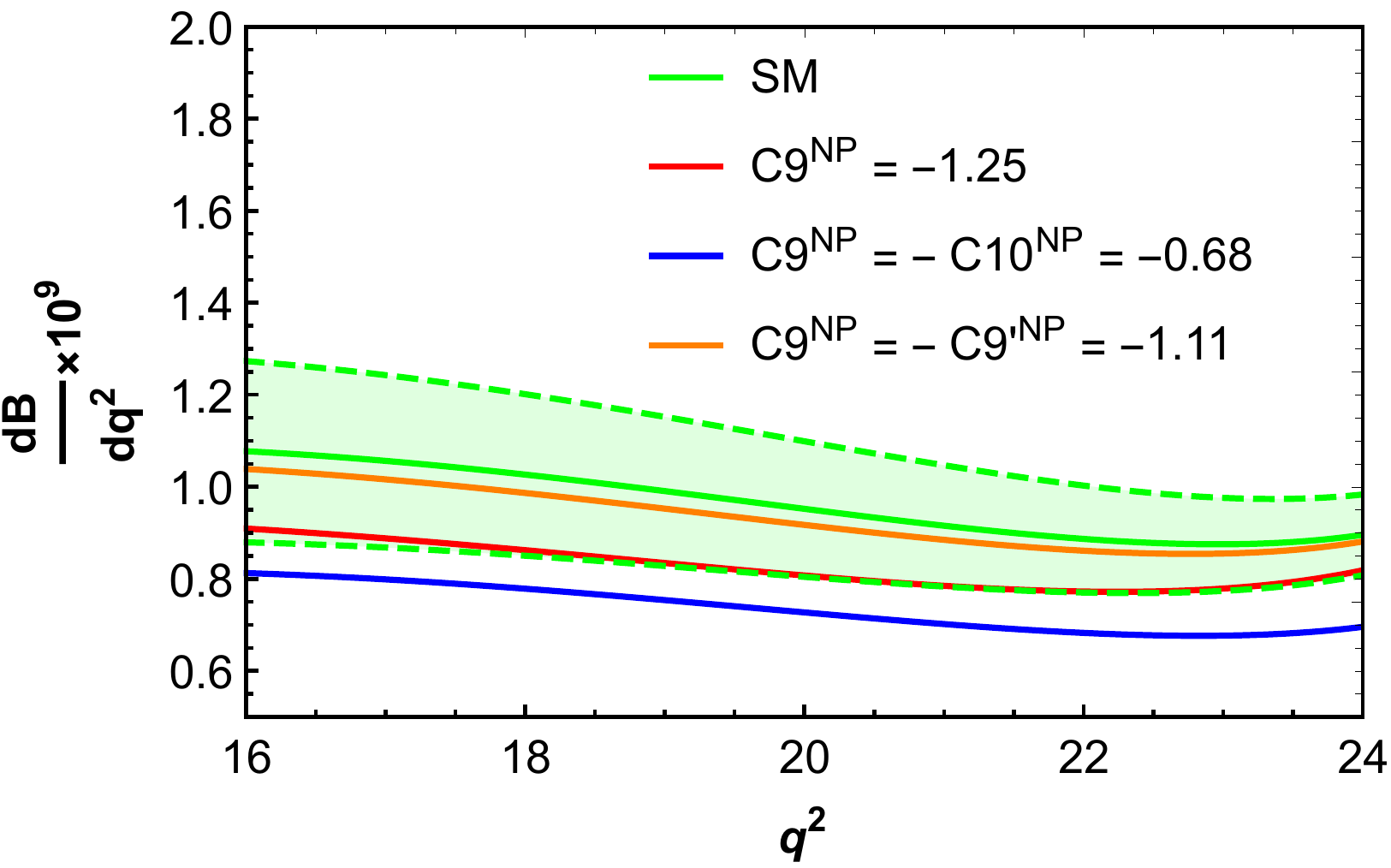}\\
\end{tabular}
\caption{Left and right panels depict the differential branching ratio, $dB/dq^2$, in low (2-6 $\mathrm{GeV}^2$) and high-$q^2$ (15.8-23 $\mathrm{GeV}^2$) regions, respectively for the double pole parameterization of the form factors. The green band corresponds to the 1$\sigma$ range of the SM prediction.  The red, blue and orange curves correspond to $dB/dq^2$ for scenarios (I), (II) and (III), respectively at the best fit values of the new physics WCs.}
\label{fig-dbr2}
\end{figure}

The predictions for various observables are obtained in the low- and high-$q^2$ regions. We choose low-$q^2$ region as 2 $\mathrm{GeV}^2$ $\leq$ $q^2$  $\leq$ 6 $\mathrm{GeV}^2$ as the dominant contribution in this region comes mainly from the diagrams where the final state photon is emitted either from  the bottom or strange quark and hence the decay is driven mainly by the $b \to s\, \mu^+ \,\mu^-$   effective Hamiltonian \cite{Alok:2010zd}. The high-$q^2$ region is chosen as 15.8 $\mathrm{GeV}^2$ $\leq$ $q^2$ $\leq$ 23 $\mathrm{GeV}^2$,  the reasons for which is explained below.

The differential branching ratio  $dB/dq^2$ in the low and high $q^2$ regions corresponding to single and double pole form-factor parametrizations for various new physics scenarios and SM are depicted in Figs.~\ref{fig-dbr1} and \ref{fig-dbr2}, respectively. From these figures, it can be seen  that none of the new physics scenarios can provide large deviation from the SM prediction. This can be further seen from the integrated values of the branching ratio of $B_s \to \mu^+ \, \mu^-\, \gamma$. The integrated values of $\mathcal{B}(B_s \to \mu^+ \, \mu^-\, \gamma)$  in the SM and in the various new physics scenarios for single and double pole parametrization of the form factors are given in Table.~\ref{tab2}. Here we have added uncertainties in the form-factors, CKM matrix elements \cite{pdg} and the contribution of the light vector meson $\phi$ (in the low $q^2$ region) \cite{Khodjamirian:2010vf}
in quadrature.  From the table, it can be seen  that the predictions for all new physics scenarios are consistent with the SM. This conclusion is independent of the choice of form factor parametrization considered in this work. Therefore the decay $B_s \to \mu^+ \, \mu^-\, \gamma$ is expected to be observed with a branching ratio close to its SM prediction.

\begin{table}[h]
  \begin{center}
\begin{tabular}{|c|c|c|c|c|}
\hline
 Scenario & \multicolumn{2}{|c|}{$\mathcal{B}$: Double Pole} & \multicolumn{2}{|c|}{$\mathcal{B}$: Single Pole} \\ \hline
  & Low $(\times 10^{-10})$ & High $(\times 10^{-10})$ & Low $(\times 10^{-10})$& High $(\times 10^{-10})$ \\  
 \hline
SM &$\,1.50 \pm 0.29 \, $  & $\,2.44 \pm 0.36\, $ & $\, 1.80 \pm 0.16\, $ & $\,4.97 \pm 0.78\,$\\
\hline 
(I) $C_9^{\mu \mu}$ (NP) & $\, 1.89 \pm 0.37 \, $   & $\, 2.07 \pm 0.28 \,$ & $\,2.24 \pm 0.22\,$ & $\,4.05 \pm 0.61\,$  \\
  \hline
(II) $C_9^{\mu \mu}$ (NP) = - $C_{10}^{\mu \mu}$ (NP) & $\,1.51 \pm 0.33 \,$ & $\, 1.67 \pm 0.26 \,$ & $1.71 \pm 0.18\,$ & $\,3.39 \pm 0.57\,$ \\
   \hline
(III) $C_9^{\mu \mu}$ (NP) = - $C_{9}^{'\mu \mu}$ (NP) &  $\, 1.71 \pm 0.34 \,$ &  $\, 2.35 \pm 0.36 \,$ & $\,2.07 \pm 0.20\,$ & $\,4.52 \pm 0.74\,$ \\
    \hline
    \end{tabular}
        \caption{Integrated values of the branching ratio $\mathcal{B}(B_s \rightarrow \mu^+ \mu^- \gamma)$ in the low $q^2$ (2-6 $\mathrm{GeV}^2$) and high $q^2$ (15.8-23 $\mathrm{GeV}^2$) region. }
            \label{tab2}
  \end{center}
\end{table}

\begin{figure}[t] 
\centering
\begin{tabular}{cc}
\includegraphics[width=82mm]{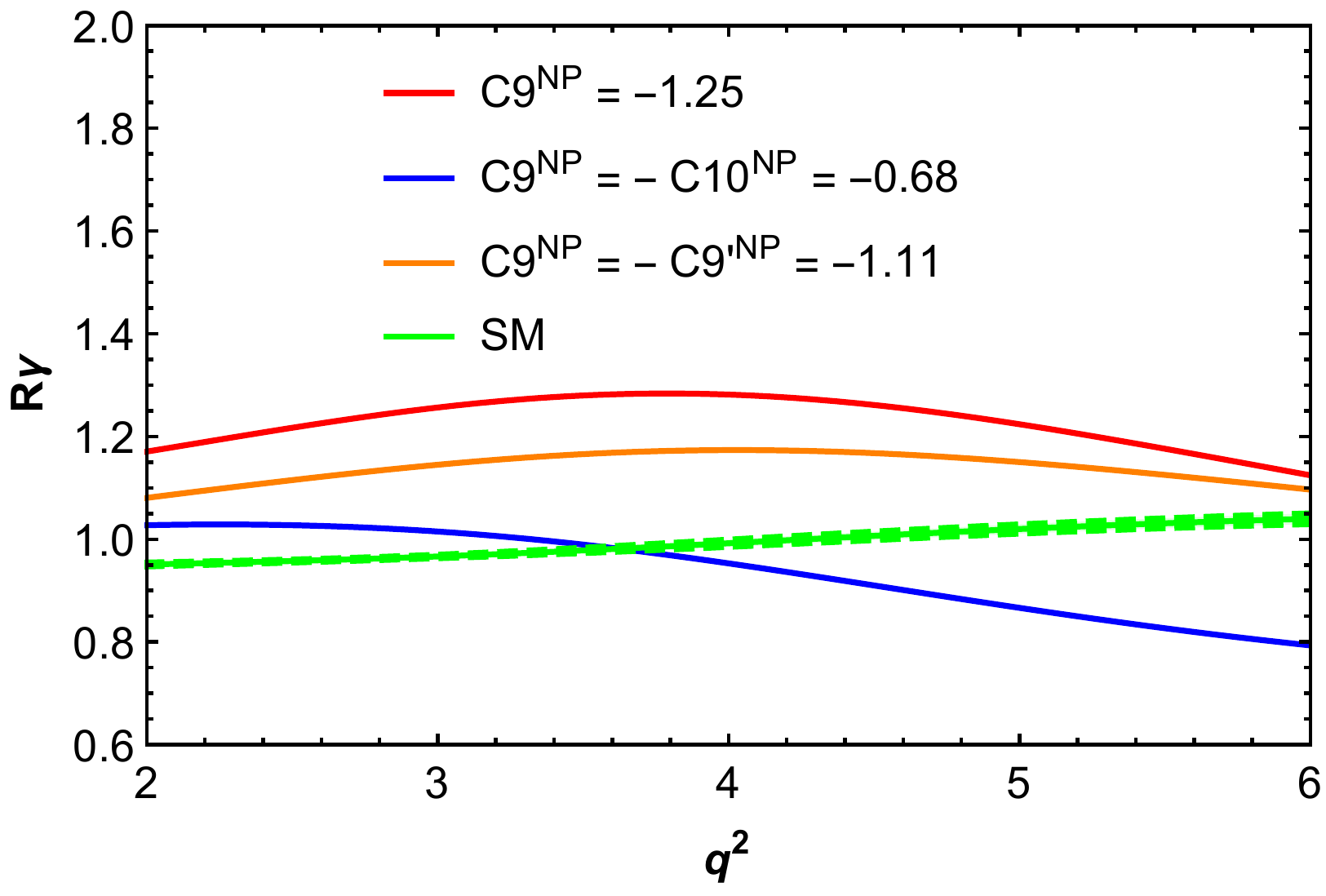}&
\includegraphics[width=82mm]{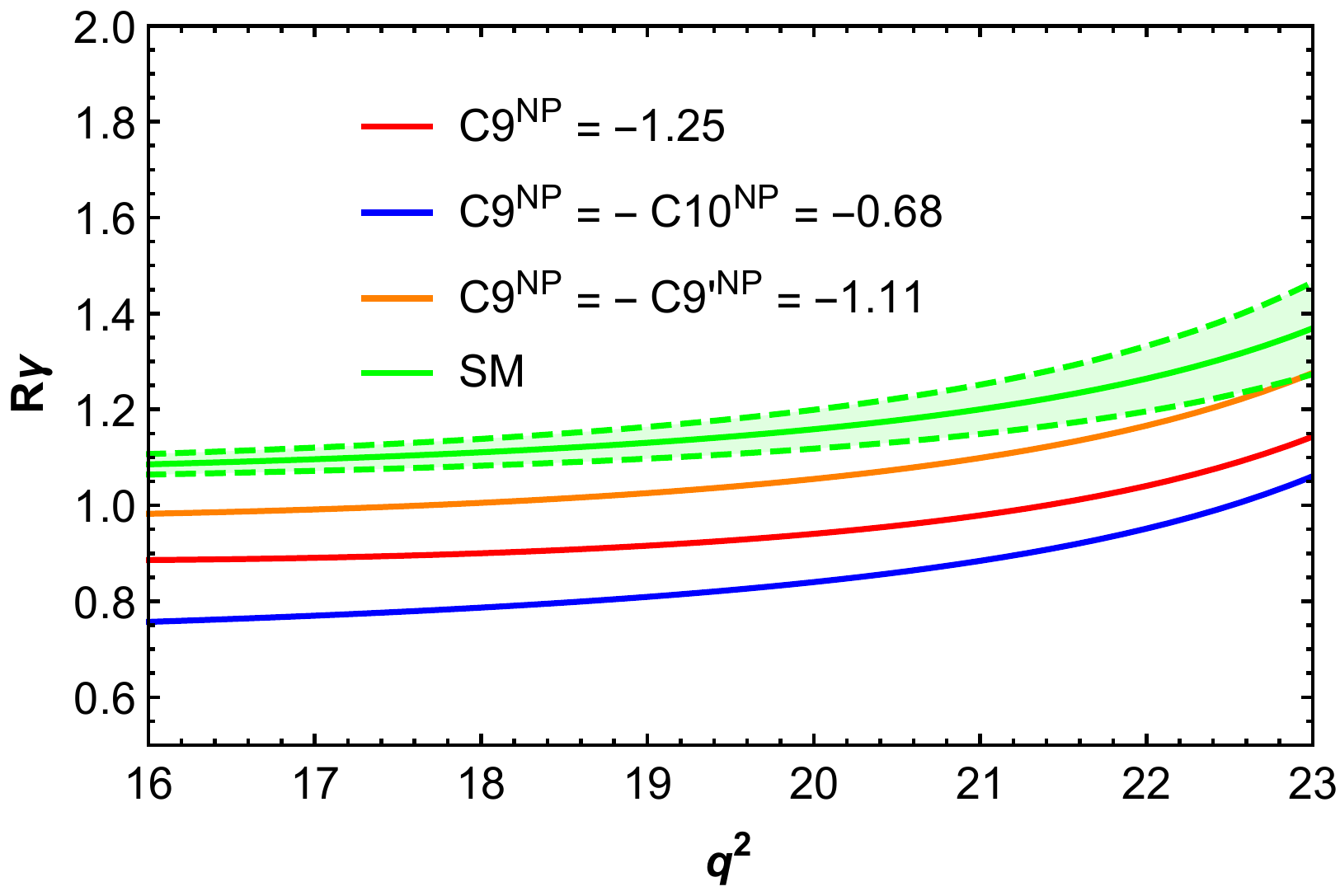}\\
\end{tabular}
\caption{Left and right panel depicts  $R_{\gamma}(q^2)$ in the low (2-6 $\mathrm{GeV}^2$) and high-$q^2$  (15.8-23 $\mathrm{GeV}^2$) regions, respectively for the single pole parametrization of the form factors. The green band corresponds to the 1$\sigma$ range of the SM prediction. The red, blue and orange curves correspond to $R_{\gamma}$ for scenarios (I), (II) and (III), respectively at the best fit values of the new physics WCs.}
\label{fig-rs}
\end{figure}

\begin{figure}[t] 
\centering
\begin{tabular}{cc}
\includegraphics[width=82mm]{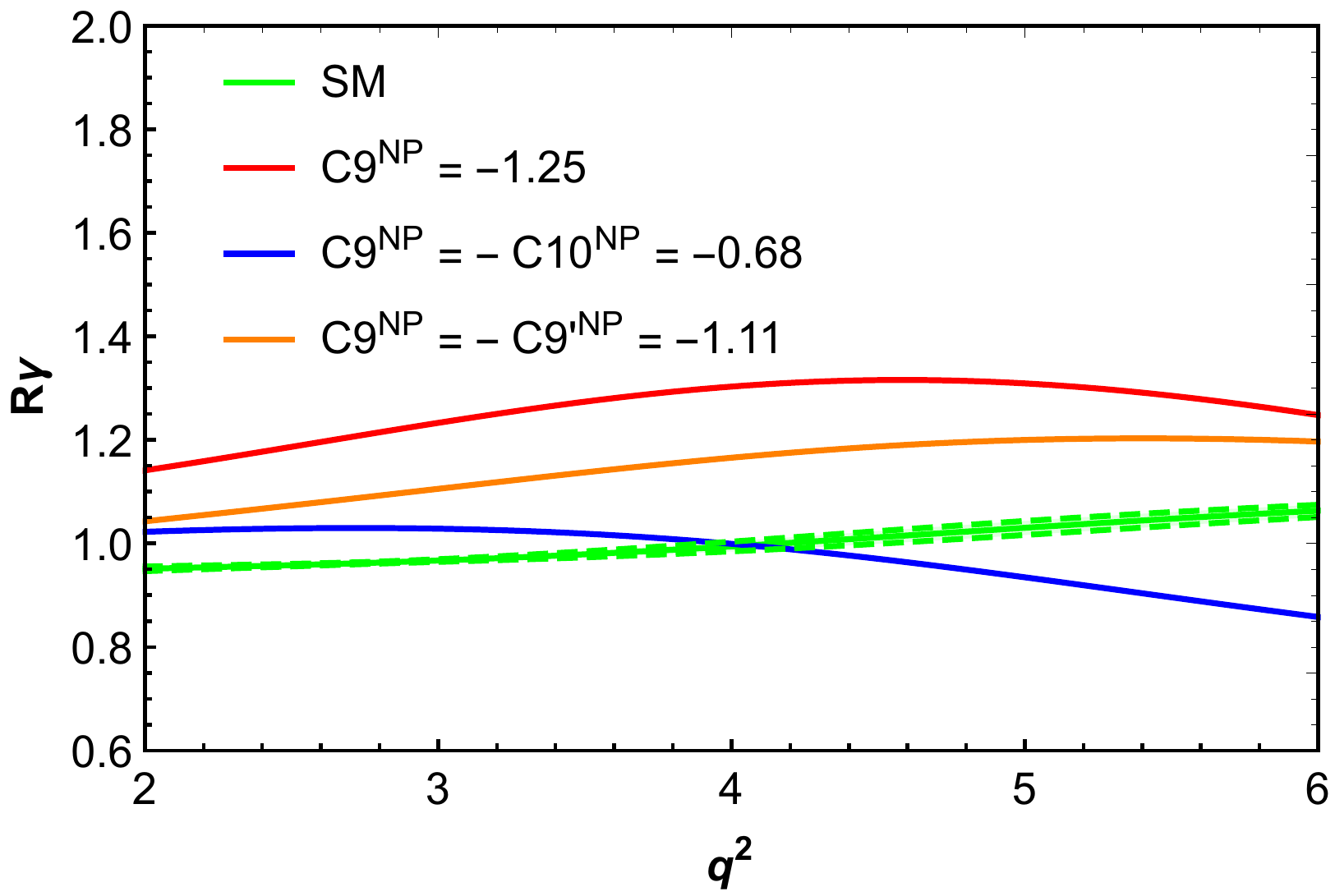}&
\includegraphics[width=82mm]{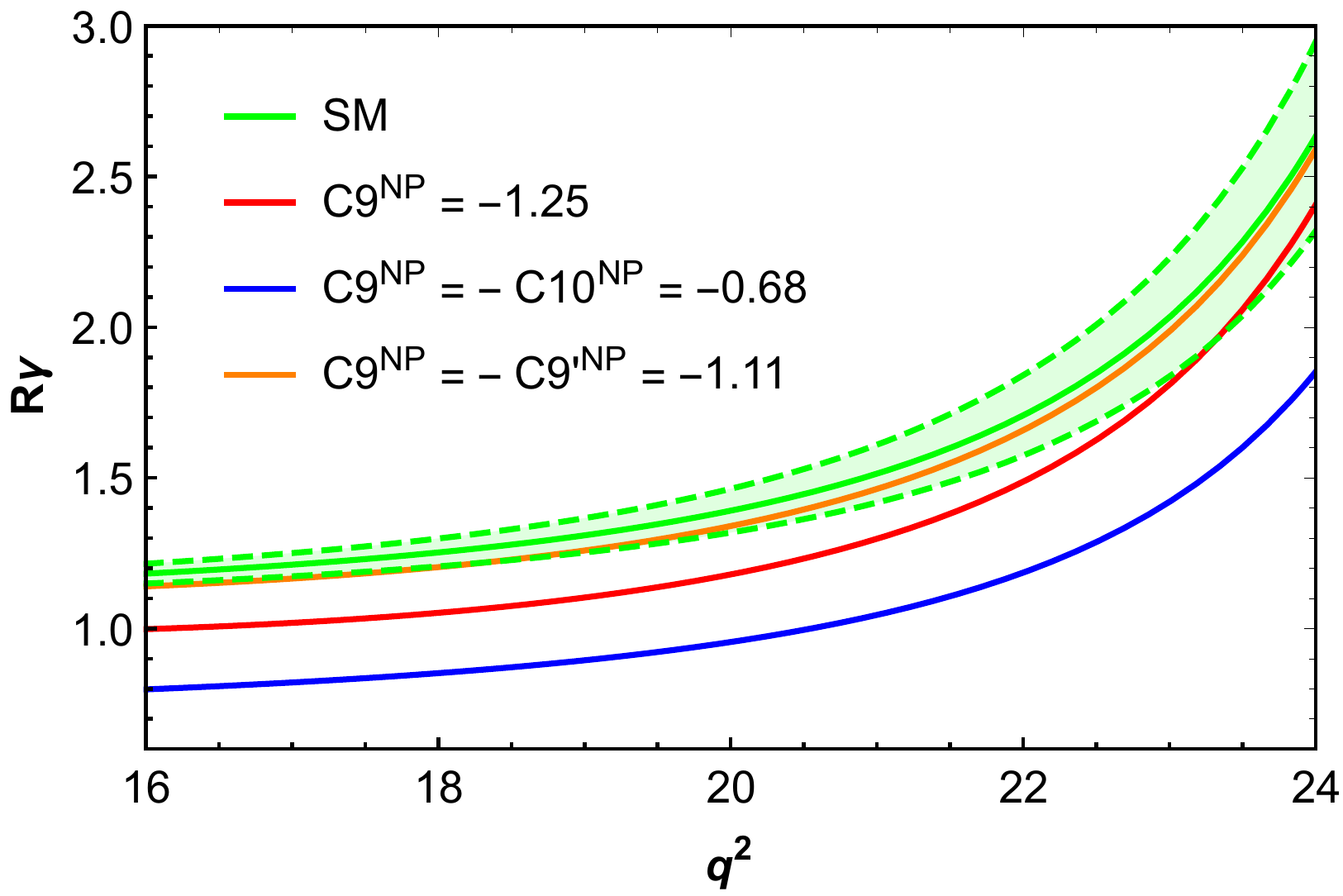}\\
\end{tabular}
\caption{Left and right panel depicts  $R_{\gamma}(q^2)$ in the low (2-6 $\mathrm{GeV}^2$) and high-$q^2$  (15.8-23 $\mathrm{GeV}^2$) regions, respectively for the double pole parametrization of the form factors. The green band corresponds to the 1$\sigma$ range of the SM prediction. The red, blue and orange curves correspond to $R_{\gamma}$ for scenarios (I), (II) and (III), respectively at the best fit values of the new physics WCs.}
\label{fig-rd}
\end{figure}

The ratio  $R_{\gamma}(q^2)$  in the low and high $q^2$ regions corresponding to single and double pole parametrizations of the form factors in the SM and for various new physics scenarios are presented in Figs.~\ref{fig-rs} and \ref{fig-rd}, respectively. It can be seen from the figures that the SM predictions of $R_{\gamma}(q^2)$ is close to $ 1$ in the entire low-$q^2$ region for both form factor parametrizations. In the high-$q^2$ region, $R_{\gamma}(q^2) \sim 1$ for $q^2<$ 18 $\rm GeV^2$. Above  $q^2 =$ 18 $\rm GeV^2$,  the value of $R_{\gamma}(q^2)$ starts to increase from unity and at the extreme end of the $q^2$ spectrum, $R_{\gamma}(q^2)$ increases upto $3.5$ for double pole parametrization.  The value of $R_\gamma$ deviates from unity mainly due to lepton mass effects in the bremsstrahlung contribution to the $B_s \rightarrow l^+ l^- \gamma$ decay rate.  At low $q^2$, the bremsstrahlung amplitude is suppressed by O$(m_l/m_{Bs})$ compared the amplitude $A^{(1)}$. At higher $q^2$ values when $q^2 \rightarrow M_{Bs}^2$, the contribution from bremsstrahlung being proportional to $m_l^2/M_{Bs}^3\,\times (M_{Bs}^4 + q^4)/(M_{Bs}^2- q^2)$,  starts to dominate the total branching ratio and hence increases $R_\gamma$ above 1. The contribution from the interference amplitude, being proportional to $m_l^2\,(M_{Bs}^2-q^2)^2$, is small compared to the bremsstrahlung contribution at large $q^2$. The dominant contribution to $R_\gamma(q^2)$ in the high $q^2$ region comes from the terms containing the Wilson coefficients $C_9$ and $C_{10}$ while the contribution from terms containing $C_{7\gamma}$ are small.

 The uncertainties in  the form-factor cancels up to a large extent in the ratio $R_{\gamma}$. However the cancellations in these uncertainties become worse with increase in   $q^2$, in particular in the extreme high $q^2$ region \cite{Guadagnoli:2017quo}.   Hence the uncertainties in $R_{\gamma}$ increases with increase in $q^2$ in the high-$q^2$ region. This is true for both single and double pole parametrization of form-factors with uncertainty being larger in the case of double pole parametrization. For this reason we truncate the high-$q^2$ region at 23 $\rm GeV^2$.

It can be seen from the left panel of Fig.~\ref{fig-rs} that $R_{\gamma}(q^2)$, in the low-$q^2$ region,  corresponding to the best fit values of the WCs for scenarios (I) and (III) lie well beyond the SM band. For scenario (III), the deviation from SM band is more prominent for $q^2 \sim$ 4.5 $\rm GeV^2$ - 6 $\rm GeV^2$. In the high-$q^2$ region, as can be seen from the right panel of Fig.~\ref{fig-rs}, $R_{\gamma}(q^2)$ curves for scenarios (I) and (II) lie well outside the SM band, the maximum deviation being for scenario (II). The deviation is relatively less for scenario (III). This is in agreement with the findings of \cite{Guadagnoli:2017quo} where, using the single pole approximation for the form factors, it  was shown that the curve $R_{\gamma}(q^2)$ in the high-$q^2$ region for scenario (II) falls well outside the SM range of $R_{\gamma}(q^2)$. 

The predictions for $R_{\gamma}(q^2)$ using double pole parametrization of the form factors are given in Fig.~\ref{fig-rd}. From the left panel of the figure, one can infer that the results, in the low-$q^2$ region, are almost the same as that of the single pole case. However, in the high-$q^2$ range, the results differ.  $R_{\gamma}(q^2)$  curve for scenario (III) lies within the SM band. The $R_{\gamma}(q^2)$  curves for scenarios (I) and (II) still lie outside the SM band but the deviation is less as compared to that of the single pole case.

\begin{table}[h]
  \begin{center}
\begin{tabular}{|c|c|c|c|c|}
\hline
 Scenario & \multicolumn{2}{|c|}{$R_\gamma$: Double Pole} & \multicolumn{2}{|c|}{$R_\gamma$: Single Pole} \\ \hline
  & Low & High & Low & High \\  
 \hline
SM &$\,0.99 \pm 0.01 \, $  & $\,1.37 \pm 0.07\, $ & $\, 0.99 \pm 0.01\, $ & $\,1.15 \pm 0.02\,$\\
\hline 
(I) $C_9^{\mu \mu}$ (NP) & $\, 1.24 \pm 0.05\, $   & $\, 1.17 \pm 0.08 \,$ & $\,1.23 \pm 0.05\,$ & $\,0.94 \pm 0.04\,$  \\
  \hline
(II) $C_9^{\mu \mu}$ (NP) = - $C_{10}^{\mu \mu}$ (NP) & $\,0.99 \pm 0.04 \,$ & $\, 0.94 \pm 0.07 \,$ & $\,0.95 \pm 0.04\,$ & $\,0.83 \pm 0.04\,$ \\
   \hline
(III) $C_9^{\mu \mu}$ (NP) = - $C_{9}^{'\mu \mu}$ (NP) &  $\,1.13 \pm 0.04 \,$ &  $\, 1.32 \pm 0.08 \,$ & $\,1.14 \pm 0.04\,$ & $\,1.05 \pm 0.05\,$ \\
    \hline
    \end{tabular}
        \caption{Integrated values of $R_\gamma$ in the low $q^2$ (2 $\mathrm{GeV}^2$ $\leq q^2 \leq$   6 $\mathrm{GeV}^2$) and high $q^2$ (15.8 $\mathrm{GeV}^2$ $\leq q^2 \leq$   23 $\mathrm{GeV}^2$) region.}
            \label{table-rg}
  \end{center}
\end{table}

 The integrated values of $R_{\gamma}(q^2)$, $R_{\gamma}$, for single and double pole scenarios are listed in Table.~\ref{table-rg}. Here the uncertainties due to form factors are added in quadrature. In the low-$q^2$ region, we have included additional uncertainties related to the light vector meson $\phi$.  For new physics scenarios, the error in new physics WCs as given in Table.~\ref{table-wc}, are also included.

 For the single pole parametrization of the form factors, the predictions for $R_{\gamma}$ in the low-$q^2$ region for scenarios (I) and (III) deviates from its SM prediction by 4$\sigma$ and 3$\sigma$, respectively. $R_{\gamma}$ for scenario (II) is consistent with the SM. 
For the double pole form factor parametrization, $R_{\gamma}$  for scenario (II) is consistent with the SM while $R_{\gamma}$  for scenarios (I) and (III) deviates from the SM at the level of 4.1$\sigma$ and 2.8$\sigma$, respectively.  
Thus we see that the conclusions are almost independent of the choice of form factor parametrization in the low-$q^2$ region.

 For the single pole case, the predictions for $R_{\gamma}$ in the high-$q^2$ region for scenarios (I), (II) and (III) deviates from its SM prediction by 3.5$\sigma$, 5.3$\sigma$ and 1.4$\sigma$, respectively. 
For the double pole case, $R_{\gamma}$  for scenario (III) is consistent with the SM. $R_{\gamma}$  for scenarios (I) and (II) deviates from the SM at the level of 1.3$\sigma$ and 3$\sigma$, respectively. 
Thus the conclusions in the high-$q^2$ region rely heavily on the choice of form factor parametrization.

We now consider NP effects in FB asymmetry of muons in $B_s \to \mu^+ \, \mu^-\, \gamma$. For single pole parametrization, the SM predictions for the integrated values  of $A_{FB}(q^2)$,  $\langle A_{FB}(q^2) \rangle$,  in the low and high $q^2$ regions are $0.48 \pm 0.05$ and $-0.58\pm 0.03$, respectively. 
For scenarios (I), (II) and (III), the prediction for  $\langle A_{FB}(q^2) \rangle$, in the low-$q^2$ region, are $(0.58 \pm 0.02)$, $(0.55 \pm 0.03)$ and $(0.42 \pm 0.05)$, respectively. In the high-$q^2$ region the predictions for NP scenarios (I), (II) and (III) are $(-0.42 \pm 0.05)$, $(-0.56 \pm 0.04)$ and $(-0.64 \pm 0.06)$, respectively.  Thus we see that the predictions for all NP scenarios  are consistent with the SM value. These conclusions remain the same for double pole parametrization as well.

\section{Conclusions}

The measurements of several observables in the decays induced by the  quark level transition  $b \to s \mu^+ \mu^-$ do not agree with the predictions of SM. These measurements could be considered as hints of  physics beyond the SM. Several new physics scenarios, all   in the form of vector and axial-vector operators, were suggested as an explanation of anomalies in the $b \to s \mu^+ \mu^-$  sector.  Therefore it is worth to consider the impact of these solutions on other related decay modes.

In this work we study new physics effects on the radiative leptonic decay of  $B_s$ meson in the light of the present $b \to s \mu^+ \mu^-$ data. We consider contributions to the $b \to s \mu^+ \mu^- \gamma $ decay from: (i) direct emission of real or virtual photons from the valence quarks of the $B_s$ meson,
(ii) emission of real photon from the $b \rightarrow s$ loop, (iii) weak annihilation, 
and (iv) bremsstrahlung from leptons in the final state.
We compute the branching ratio of $B_s \to \mu^+ \mu^- \gamma$, the ratio $R_{\gamma}$  of the differential distribution $B_s \to \mu^+ \mu^- \gamma$ and  $B_s \to e^+ e^- \gamma$  and the muon forward-backward asymmetry $A_{FB}$  in the presence of the additional new physics vector and axial-vector operators.
We consider the form factors relevant for this decay both in the single pole and modified pole parameterization and obtain predictions for these quantities for all allowed new physics solutions. 

We find that for all allowed new physics solutions, the predicted values of the branching ratio and $A_{FB}$ are consistent with the SM. However, a large deviation in $R_{\gamma}$ as compared to its SM value is allowed for some of the new physics solutions. The prediction of $R_{\gamma}$, in the low-$q^2$ region (2-6 $\rm GeV^2$), for $C_9^{\mu \mu}$ (NP) $<0$ solution deviates from its SM prediction at the level of 4$\sigma$ for both single  and modified pole parameterization of the form-factors. The conclusions in the high-$q^2$ region  (16-23 $\rm GeV^2$) rely heavily on the choice of parametrization of the form-factors.  However, both parametrizations allow 3$\sigma$ deviation in $R_{\gamma}$ for   $C_9^{\mu \mu}$ (NP) = - $C_{10}^{\mu \mu}$ (NP) $<0$ solution. Hence the measurement of $R_{\gamma}$ can be useful in identifying  the Lorentz structure of NP in $b \to s \mu^+ \mu^-$ transition.

\section{Acknowledgment}
 We are thankful to Diego Guadagnoli for useful suggestions regarding our analyses related to $R_{\gamma}$.  We also thank S. Uma Sankar, Dinesk Kumar and Jacky Kumar for useful discussions and suggestions.

\end{document}